\documentclass[proof]{WileyASNA-v1}

\articletype{Article Type}%

\received{26 April 2016}
\revised{6 June 2016}
\accepted{6 June 2016}

\begin{document}

\title{XMM2ATHENA, the H2020 project to improve XMM-Newton analysis software and prepare for Athena}

\author[1]{Natalie A. Webb}
\author[2]{Francisco J. Carrera}
\author[3]{Axel Schwope}
\author[4]{Christian Motch}
\author[5]{Jean Ballet}
\author[6]{Mike Watson}
\author[7]{Mat Page}
\author[8]{Michael Freyberg}
\author[9]{Ioannis Georgantopoulos}
\address[1]{IRAP, Universit\'e de Toulouse, CNRS, CNES, 9 avenue du Colonel Roche, 21028 Toulouse, France}
\address[2]{Instituto de F\'\i{}sica de Cantabria (CSIC-Universidad de Cantabria), Avenida de los Castros, 39005 Santander, Spain}
\address[3]{Leibniz-Institut für Astrophysik Potsdam (AIP), An der Sternwarte 16, 14482 Potsdam, Germany}
\address[4]{Universit\'e de Strasbourg, CNRS, Observatoire astronomique de Strasbourg, UMR 7550, 67000 Strasbourg, France}
\address[5]{Universit\'e Paris-Saclay, Universit\'e Paris Cit\'e, CEA, CNRS, AIM, F-91191 Gif-sur-Yvette Cedex, France}
\address[6]{Department of Physics \& Astronomy, University of Leicester, Leicester, LE1 7RH, UK}
\address[7]{Mullard Space Science Laboratory, University College London, Holbury St Mary, Dorking, Surrey RH5 6NT, UK}
\address[8]{Max-Planck-Institut f\"ur extraterrestrische Physik, Giessenbachstra{\ss}e 1, 85748 Garching, Germany}
\address[9]{IAASARS, National Observatory of Athens, I. Metaxa \& V. Pavlou, 15236, Greece}
\author[1]{Mickael Coriat}
\author[1]{Didier Barret}
\author[1]{Zoe Massida}
\author[1]{Maitrayee Gupta}
\author[1]{Hugo Tranin}
\author[1]{Erwan Quintin}
\author[2]{M. Teresa Ceballos}
\author[2]{Silvia Mateos}
\author[2]{Amalia Corral}
\author[2]{Rosa Dominguez}
\author[2]{Holger Stiele}
\author[3]{Iris Traulsen}
\author[3]{Adriana Pires}
\author[4]{Ada Nebot}
\author[4]{Laurent Michel}
\author[4]{François Xavier Pineau}
\author[4]{Jere Kuutila}
\author[4]{Pierre Maggi}
\author[5]{Sudip Chakroborty}
\author[6]{Keir Birchall}
\author[7]{Paul Kuin}
\author[9]{Athanassios Akylas}
\author[9]{Angel Ruiz}
\author[9]{Ektoras Pouliasis}
\author[9]{Antonis Georgakakis}

\authormark{Webb \textsc{et al}}






\abstract{{\it XMM-Newton}, a {\it European Space Agency} observatory, has been observing the X-ray, ultra-violet and optical sky for 23 years. During this time, astronomy has evolved from mainly studying single sources to populations and from a single wavelength, to multi-wavelength/messenger data. We are also moving into an era of time domain astronomy. New software and methods are required to accompany evolving astronomy and prepare for the next generation X-ray observatory, {\it Athena}.

Here we present {\it XMM2ATHENA}, a programme funded by the European Union’s Horizon 2020 research and innovation programme. {\it XMM2ATHENA} builds on foundations laid by the {\it XMM-Newton Survey Science Centre} ({\it XMM-SSC}), including key members of this consortium and the {\it Athena} Science ground segment, along with members of the X-ray community. The project is developing and testing new methods and software to allow the community to follow the X-ray transient sky in quasi-real time, identify multi-wavelength/ messenger counterparts of {\it XMM-Newton} sources and determine their nature using machine learning. We detail here the first milestone delivery of the project, a new online, sensitivity estimator. We also outline other products, including the forthcoming innovative stacking procedure and detection algorithms to detect the faintest sources.  These tools will then be adapted for {\it Athena} and the newly detected/ identified sources will enhance preparation for observing the {\it Athena} X-ray sky.}



\maketitle


\section{Introduction}\label{sec:intro}

Observing the sky in X-rays allows us to detect the hottest and most energetic
phenomena in the Universe. X-rays allow us to detect matter being accreted onto
massive black holes in the centres of galaxies or onto stellar mass black holes in binaries with other stars (X-ray binaries). They allow us to identify stellar flares from active stars, supernova remnants, neutron stars, white dwarfs, galaxy clusters and even aurora on planets or comets. {\it XMM-Newton}, a {\it European Space Agency} X-ray observatory \citep{jans01} and the largest X-ray telescope to be launched to date, has been observing the X-ray, ultra-violet and optical sky for 23 years.

In order to exploit the {\it XMM-Newton} data, the {\it XMM-Newton Survey Science Centre}\footnote{\url{http://xmmssc.irap.omp.eu/}} \citep[XMM-SSC, ][]{wats01}, a consortium of ten European institutes in collaboration with the {\it XMM-Newton Science Operations Centre} (SOC), have put together a suite of software called the {\it Science Analysis System} \cite[SAS,][]{sas04}, that can be used to reduce and analyse the {\it XMM-Newton} data. A dedicated pipeline has also been developed to perform standardised routine  processing of the {\it XMM-Newton} science data.  These products are used to generate catalogues of X-ray and optical/UV sources \citep{webb20,traul20,page12}.

However, over the last 23 years, astronomy has evolved towards the study of populations rather than individual sources and to use multi-wavelength and multi-messenger data to understand the X-ray sources. We are also moving into an era of time domain astronomy, which requires operating observatories differently in order to catch outbursts from X-ray binary systems, from tidal disruption events (TDE) occurring as a massive black hole tears apart a passing star, from supernovae, from stellar flares, etc and follow them up rapidly at other wavelengths or with multi-messenger observations. New software and methods need to be put into place to accompany this new astronomy. It will be even more important in the next decade, when the next generation X-ray observatory, {\it Athena} \citep{nand13}, will be launched. Here we describe a programme {\it XMM2ATHENA}\footnote{\url{http://xmm-ssc.irap.omp.eu/xmm2athena/}}, financed through {\it Horizon 2020}, which is a European Union Framework Programme for Research and Innovation. {\it XMM2ATHENA} runs for three years, from April 2021 to March 2024 and builds on work currently done in the XMM-SSC and aims to develop new methodology to exploit the current {\it XMM-Newton} data and further develop it to fully exploit the {\it Athena} data in conjunction with multi-wavelength and multi-messenger data. We detail the first milestone product of {\it XMM2ATHENA}, a sensitivity estimator for {\it XMM-Newton} data, as well as other new tools and products already released, namely new machine learning tasks to classify both the X-ray and optical/UV sources, new spectral fitting products, enhanced algorithms to identify multi-wavelength and multi-messenger counterparts, and new time domain algorithms and summarise the development to come with regards to improved methods for source detection in stacked data, photometric redshifts and physically motivated spectral models for the best spectra. All of this information will be contained in a forthcoming catalogue version, 5XMM.

The {\it XMM2ATHENA} project also has a strong outreach aspect, with the aim of raising the profile of both {\it XMM-Newton} and {\it Athena} in the astronomical community as well as using the two observatories as a stepping stone for communicating with and educating the public.  This is done through a dedicated web page (available in four languages) along with Twitter and Facebook accounts. We have participated in national and international activities including Science
Week, European researcher’s Night, Space Week, International Day of Women and
Girls in Science and Women's Day and held online competitions and events. We have also produced a wide range of public outreach material, again available on our webpages \url{http://xmm-ssc.irap.omp.eu/xmm2athena/}.

\section{Sensitivity estimator}
\label{sec:flix}

We have developed a new sensitivity estimator based on the {\it Flux Limits from Images from XMM-Newton} (FLIX) tool. It can be used to estimate either the {\it XMM-Newton} X-ray flux or an estimate of the sensitivity of the EPIC (European Photon Imaging Cameras) detector(s) at a given point in the sky, within an extraction radius defined by the user. This is of particular interest for extended regions of interest or very bright sources. Once the parameters have been provided by the user, the tool scans the 4XMM catalogue and lists the nearest sources. HTML links are provided to access the source nature and the distance from the position of interest. The latest version of the catalogue, 4XMM-DR12 at the time of writing, is ingested into the sensitivity estimator. A single position can be entered or for multiple sources a source file can be provided. The user can choose the detection likelihood threshold for upper-limit estimation as well as the wavebands to investigate.

The sensitivity estimate is determined empirically using the algorithm described in \cite{carr07}. These estimates use the detector mask to find which pixels in the image are valid, the exposure map data to find the exposure time, and the background map to determine the fitted background level. Should a source be present at the indicated position, the flux is calculated by summing the counts inside the user defined radius, using the background level from the background map data and an approximation to the point-spread function in the calibration files. Energy conversion factors are those used in the 4XMM catalogue \citep{webb20}.

\begin{figure*}
\includegraphics[width=18cm,height=130mm]{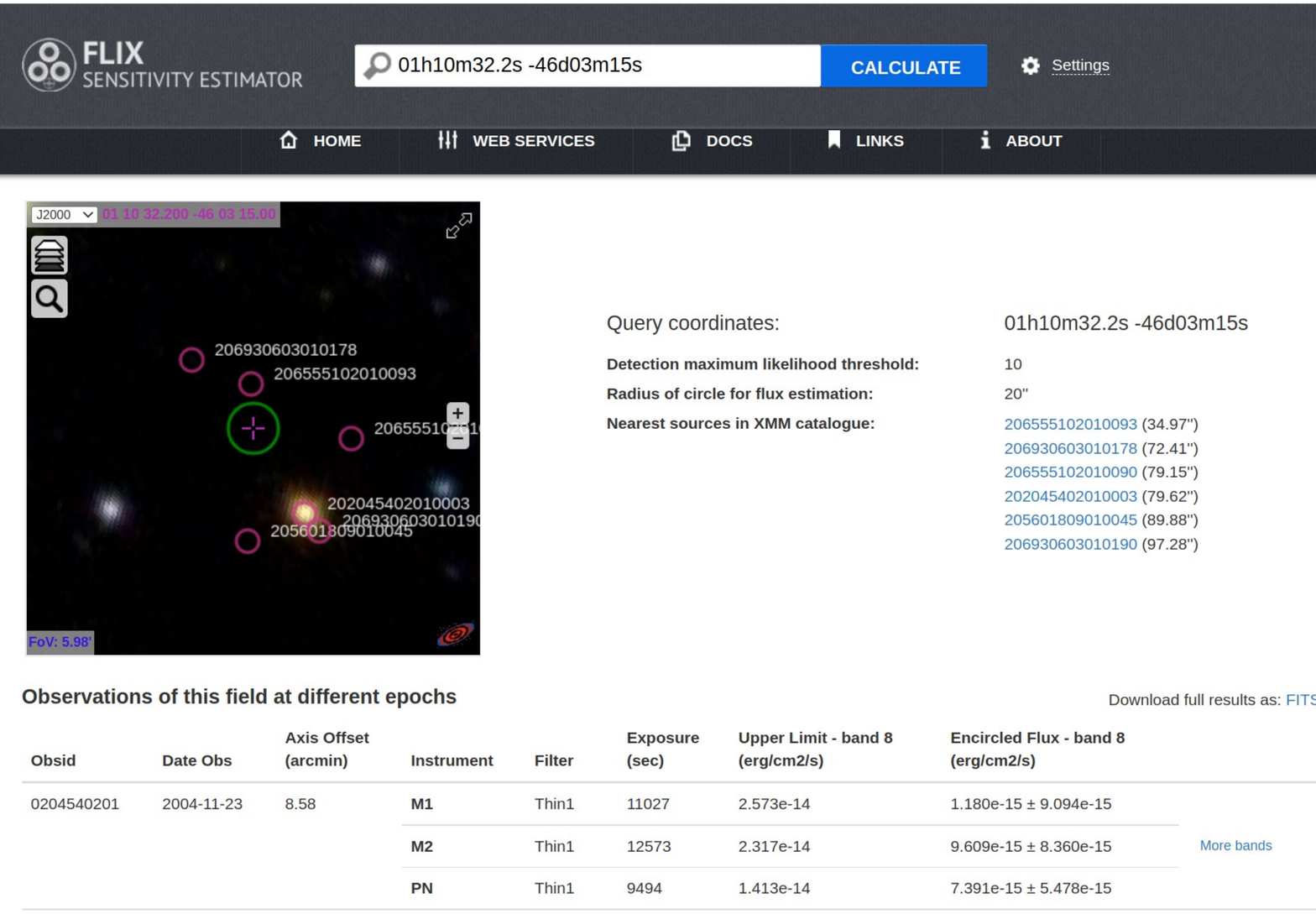}
\caption{Extract of the output from the FLIX sensitivity estimator. The coordinates of interest can be entered in the box at the top and the settings modified using the corresponding button. The three band colour image is plotted with the region of interest shown by the the cross and the green circle. Neighbouring sources are indicated with the pink circles and are included on the right with a link to their parameters, as given in the 4XMM catalogue. Images taken in other wavelengths can be chosen instead, using the stack button on the observation, thanks to the AladinLite functionality. Also indicated are the source identification number from 4XMM. Towards the bottom of the page, information is provided for every observation in which the requested region of interest was observed. This can also be retrieved in a FITS file, using the link towards the bottom right.
  }
\label{fig:FLIXoutput}
\end{figure*}

Results are returned as a web page which contains a HTML table with an entry per celestial position, for each detector and each observation concerned by the region. A FITS file is also produced, with a binary table containing one row per observation which contains useful information, notably for different energy bands. This may be downloaded as soon as the HTML output is complete, and examined with utilities such as the {\it ftool} FV \citep{blac95} or {\it TOPCAT} \citep{tayl05}. An example of the output webpage is provided in Figure~\ref{fig:FLIXoutput}.

The interface to FLIX has been modernised and harmonised with the 4XMM catalogue server interface, which is intuitive (but documentation is nonetheless provided with the tool) and better adapted for long term maintenance. It can be accessed via 
\url{http://flix.irap.omp.eu/}. The tool is being continually developed and upcoming features include integrating the stacked catalogue for deeper upper limits using a rapid Poisson estimator or through the slower Cash based approach, improving the simple aperture photometry currently used, providing a script access to FLIX and redeveloping the code in a more modern language (python), to improve portability to {\it Athena}.

FLIX is complementary to the ESA services {\it The HIgh-energy LIght-curve GeneraTor} \citep[HILIGT\footnote{http://xmmuls.esac.esa.int/hiligt/}][]{saxt22,koni22} and {\it RapidXMM} \citep{ruiz22}.  HILIGT provides fluxes and upper limits for a variety of high energy missions, including the {\it XMM-Newton} pointed EPIC data. This is done by calculating the count rate from the images and converting them to fluxes. This provides a very good approximation to the flux, but FLIX provides the sensitivity estimate using methodology that is in line with that used for the {\it XMM-Newton} catalogue fluxes, making it more accurate for the {\it XMM-Newton} data with respect to the fluxes provided by HILIGT (up to 50\% difference between FLIX and HILIGT in some cases) and RapidXMM. However, FLIX does not allow the user to query a wide range of high energy misions in the same way as HILIGT. Further, FLIX allows the user to select the size of the region for the sensitivity estimate, which is not proposed by HILIGT. This is particularly important when considering extended sources/regions of the sky, or very bright sources. Note that almost 10\% of all {\it XMM-Newton} detections are extended \cite{webb20}. {\it RapidXMM} provides upper limit estimates based on aperture photometry in a similar way to HILIGT, but pre-calculates these values with a resolution of $\sim$4" and stores them in a database, so it is much quicker to access these values than for those calculated on the fly by FLIX or HILIGT. However the estimates may not be centred exactly on the position of interest.

\section{Source Classification}\label{sec:classification}

To identify the X-ray sources, and also the optical/UV sources detected with the {\it Optical Monitor} \cite[OM,][]{maso01} on board {\it XMM-Newton}, we developed a probabilistic method, revisiting the naive Bayes classification algorithm \citep{tran22}. This approach was motivated by its intuitive nature, as an extension of the rough classification rules used in simplistic decision trees. It also delivers transparency for a particular classification. To undertake the classification of the X-ray sources we used selected columns of data from the 4XMM-DR10 version of the {\it XMM-Newton} catalogue \citep{webb20} pertaining to Galactic latitude, the ratio of the maximum to minimum flux for a source and spectral information, namely using the four hardness ratios given in the catalogue, (where HR$_{band\ i}$ = (Flux$_{band\ i+1}$ -Flux$_{band\ i}$)/(Flux$_{band\ i+1}$ + Flux$_{band\ i}$). The five energy bands are 1: 0.2-0.5 keV, 2: 0.5-1.0 keV, 3: 1.0-2.0 keV, 4: 2.0-4.5 keV and 5: 4.5-12.0 keV. We augmented the spectral information by fitting a power law and an APEC\footnote{\url{http://atomdb.org/}} model to the 0.2-12.0 keV spectrum where possible or to the fluxes in the five energy bands provided in the catalogue and used the spectral slope and temperature information from these fits. We also augmented the catalogue with multi-wavelength counterparts, using NWAY \citep{salv18} from {\it Gaia} \citep{bail21} and the GLADE catalogue \citep{daly16}, as well as a number of other catalogues \citep{tran22}. This then allowed us to use X-ray to r-band flux ratios and the X-ray to W1-band flux ratios, as well as the distance to the centre of a host galaxy and the proper motion of a {\it Gaia} association. We generated a sample of 25160 previously identified sources, broken into subgroups of AGN, stars, X-ray binaries and cataclysmic variables. We estimated the probability density for each property cited above for the different source types, which were used to compute the likelihoods to classify the sources. When a property value was missing, and if the property was not missing at random, the likelihood was replaced by the probability that a source of that class had a missing value. Then the Bayes rule was used taking into account each property for each source type, to attribute the probability pertaining to the nature of the source. An outlier class was also used to identify rare sources of other types. The performance of the algorithm was validated with a test sample. Excellent results for the precision were retrieved, 97.2\% for AGN, 98.9\% for stars,  93.7\% for X-ray binaries and  84.6\% for cataclysmic variables \citep{tran22}.  The classifications for the 4XMM-DR10 version of the catalogue are available at : \url{https://vizier.cds.unistra.fr/viz-bin/VizieR-3?-source=J/A%2bA/657/A138/table7}. As of version 5XMM of the {\it XMM-Newton} catalogue, the classification will be included in the catalogue, see Section~\ref{sec:to_come}.  This algorithm can also be used to classify X-ray sources from other observatories \citep{tran22}.

Further, in order to improve the sample sizes for the learning sample and thus improve both the $precision$ (value calculated from running the classification code on a sample of known sources, where the precision is the percentage of the group of sources identified as being from a source class, that are indeed of that source class) and the $recall$ (the fraction of each source class retrieved by the algorithm), a citizen science project has been set up called {\it \underline{Cla}ssification of \underline{X}-ray \underline{So}urces for \underline{N}ovices} (CLAXSON)\footnote{\url{http://xmm-ssc.irap.omp.eu/claxson/index.php}}. This project trains users to identify the types of sources classified in \cite{tran22} and then once they reach a good level of accuracy, can classify new sources. A source is only considered as identified once at least 20 people have classified the source, with converging identifications. This platform can also be used for outreach and educative purposes.

\section{Spectral fitting}
\label{sec:spectral_fits}

Of the 895415 detections in the 4XMM-DR11 catalogue, 319565 have a sufficient number of counts to trigger automatic spectral extraction in the processing pipeline. Within the {\it XMM2ATHENA} project we aim to fit all the spectra with both simple and physically motivated models. Before performing spectral fits, we applied a quality filter to choose the good quality spectra. This filter required detections with background spectra with a positive number of counts and source+background spectra with more counts than their corresponding background spectra, scaled to the source extraction area. 311139 spectra passed these criteria. We also took into account the quality of the fits, defining a {\it good fit sample} of 232350 detections, for which the goodness of fit to both the background and source+background was $\geq$0.01. Using a sophisticated Bayesian fitting method \citep[Bayesian X-ray Analysis, BXA][]{buch14} with a thorough exploration of the parameter space, we fitted these spectra with an absorbed power law. 
The fit results are provided in a FITS table, with one row per detection\footnote{\url{http://xmm-ssc.irap.omp.eu/xmm2athena/catalogues/}}. It includes estimates of the central value along with the 90\% confidence intervals, the statistical quality of the background and source+background fits, and a number of flags, both from the quality filtering described above and that given in the 4XMM-DR11 catalogue.

We performed several quality checks on the results. 90\% of the fluxes obtained from fitting corresponded well with the fluxes provided in 4XMM-DR11, with the worst deviations coming from spectra that had best-fit parameters indicating that they were quite different from the simple power law of $\Gamma$=1.7 and an absorption of 3 $\times$ 10$^{20}$ cm$^{-2}$ that is used to derive the flux in the catalogue. We also provide fits for repeated observations of the same source to get better constraints
on its spectral shape. Of the 358809 unique sources in the catalogue of stacked data,
4XMM-DR11s, 33061 have at least two detections with spectra. We merge spectra from the same instrument (pn and MOS1/2, separately) from these repeated observations, using the SAS tool epicspeccombine (\url{https://xmm-tools.cosmos.esa.int/external/sas/current/doc/epicspeccombine/index.html}). Not all spectra are merged for each source: we start with the spectrum with the highest signal-to-noise (SNR) ratio and keep adding spectra with decreasing SNR until the overall SNR no longer improves significantly. These fits are also available on the {\it XMM2ATHENA}  webpage\footnote{\url{http://xmm-ssc.irap.omp.eu/xmm2athena/catalogues/}}.

We also provide fits for the faintest sources, for which no spectrum has been extracted, but fluxes in the five energy bands of the catalogue are available. Finally, using the source identification and the photometric redshifts, we will also provide physically motivated fits to sources with a reliable classification.

\section{Cross-matching multi-wavelength/messenger counterparts}
\label{sec:xmatch}

Using a statistical framework for multi-catalogue cross-correlation and cross-identification previously developed \citep{pine17}, based on both astrometric and photometric data, we are able to identify multi-wavelength and multi-messenger counterparts for the X-ray sources. We have extended existing methodology to new catalogues of interest and generated spectral energy distributions (SEDs). Using the SEDs we have automatically classified the sources into extragalactic and Galactic sources and we will tailor this method to the stacked catalogue and focus on some regions of sky with the deepest, best quality  multi-wavelength data available to improve the identifications.

\section{Variable sources}
\label{sec:variability}

Currently, the 4XMM catalogue provides information on variability during a single observation. Variability between observations can be probed with the stacked catalogue. However, to extend the duration over which variability can be detected and to increase the number of observations, we have taken six X-ray catalogues in addition to the {\it XMM-Newton} pointed data. These include The {\it Chandra Source Catalogue} \cite[CSC 2.0,][]{evan20b}, {\it The Swift X-ray Point Source catalogue} \cite[2SXPS,][]{evan20}, {\it The Rosat All Sky survey} \citep[2RXS,][]{boll16}, the {\it ROSAT pointed survey} \citep[WGACAT,][]{whit94}, The {\it XMM-Newton slew survey} \citep[XMMSL2,][]{saxt08} and the early release {\it eRosita} data \citep[eFEDS][]{brun22}, which we further augmented with {\it XMM-Newton} upper limits using RapidXMM. To do this we developed an algorithm based on the matching method described in \cite{salv18}. We carried out in depth analysis to determine the optimal X-ray band(s) to be compared and the spectral model to be used to determine the fluxes, in order to create reliable comparisons. We chose  the range 0.1-12.0 keV, extrapolating the flux for each observatory using our chosen spectral model, a power law with $\Gamma = 1.7$ and n$_H$ = 3 $\times$ 10$^{20}$ cm$^{-2}$. We then determined the ideal criteria for determining variability and carried out a pilot study on two months of data to determine the number and type of alerts expected. We then extended this to search for spectral variability. Finally, we also incorporated the optical/UV data from the co-aligned OM telescope and searched for flux and spectral variability in these longer wavelengths. We have also invested considerable effort in testing the reliability of the algorithm. The code can be used to search for variable sources in quasi-realtime and provide alerts to the community. The work is detailed in an upcoming paper, Quintin et al. (to be submitted).

To search for faint, short term variability, that may not be detected in the pipeline processing due to the fact that variability analysis is only carried out on sources with at least 100 counts, we have built on preliminary work developed \cite{past20} on the {\it EPIC-pn XMM-Newton outburst detector} (EXOD) algorithm. This code breaks the whole observation into short time windows and searches for variability in a way that is agnostic from source detection by looking for areas that are variable on the detector on a window to window basis. Improvements that have been implemented with regards to the original python code are extending the use of the algorithm to be applied to the MOS detectors and also to be able to use all modes of both the MOS and the pn detectors, which has allowed us to run the code on all of the 15000 observations in the {\it XMM-Newton} archive. We have also improved the accuracy of the source location and excluded regions on the detectors affected by very bright sources, including the readout streaks associated with these sources and that can lead to false detections. We have investigated a full range of detection limits in order to compare all three cameras and therefore improve the reliability of the detection. We have carried out preliminary tests on the improved algorithm, comparing it to the previous work done and by testing it on observations including specific source types, such as fields in which Fast Radio Bursts (FRBs) and Quasi-Periodic Eruptions (QPEs) have been detected and we have been able to confirm the adaptability of the code to these source types. We have also shown that the code is able to pick out significant variability from brighter sources. Whilst these sources are detected by the {\it XMM-Newton} pipeline, the variability has not been identified using the current algorithms in the pipeline.

\section{Other upcoming products}
\label{sec:to_come}  

Whilst the data coming from the OM telescope has already been correlated with the X-ray data, little work has been done to classify the sources and exploit the long term variability of the sources. The naive Bayes algorithm \citep{tran22} will be used to classify the optical/UV sources and in the framework of the long term variability analysis of the X-ray sources, the variability is currently being investigated, see Sec.~\ref{sec:variability}. This data will also be used, in conjunction with other optical/infra-red catalogues, to determine photometric redshifts for the X-ray sources, essential for improving the classification, and defining physically motivated models for spectral fits. This, along with the other products described above, will be provided in a new version of the catalogue, 5XMM, expected for 2025. For the current version of the catalogue, 4XMM, different catalogues are provided, notably a catalogue of detections, where the current version, 4XMM-DR12, includes 939270 X-ray detections which relate to 630347 unique X-ray sources from 12712 observations that were public by the 31st December 2021 and a stacked catalogue, 4XMM-DR12s, created from overlapping observations. Stacking results in better source parameters, higher sensitivity, and direct access to measures of long-term flux variability. 4XMM-DR12s is built from 1620 groups drawn from 9355 observations. For 5XMM we will provide a single catalogue comprising both single detections and stacked sources, using new methodology to reduce the number of free parameters, thus enabling the faintest sources to be detected. 5XMM will therefore be a simplified, but more complete catalogue as it will contain multi-wavelength counterparts, sensitivity estimates when no detection is made, long and short term variability, spectral fitting parameters, source classification and photometric redshifts, in addition to the $>$300 columns of information already provided.

\section{Conclusions}
\label{sec:conc}  

The {\it XMM2ATHENA} project will provide new software and {\it XMM-Newton} products, above and beyond those already provided by the XMM-SSC. Thanks to the involvement of the XMM-SSC in the {\it XMM2ATHENA} project, these products will continue to be developed and maintained even after the end of project in 2024. The main product will be a new simplified, but more complete catalogue of {\it XMM-Newton} detections and sources, but other software and tools have already been or will be made available, such as improved cross-matching tools, a sensitivity estimator, transient detection software, spectral fits, new classification software and photometric redshifts for the X-ray sources. Further, these will all be well tested thanks to four scientific projects designed to incorporate all of the tools and the products, in order to validate them. Once validated, the tools will be further developed to adapt them for the upcoming {\it Athena} mission, ensuring an enhanced scientific return for both ESA flagships.


\section*{Acknowledgments}

This project has received funding from the European Union’s Horizon 2020 research and innovation programme under grant agreement Number 101004168. NW, MC, DB, HT and EQ are also grateful for the support provided by the CNES. We are grateful to the anonymous referee who provided some excellent comments that allowed us to significantly improve this paper.









\bibliography{webb}



\end{document}